\documentclass[prd,floatfix,preprintnumbers]{revtex4}
\usepackage{amssymb}

\usepackage{amsmath}
\usepackage{graphicx,epsfig}
\usepackage{color}

\begin{document}

\title{Does inhomogeneous big bang nucleosynthesis produce
an inhomogeneous element distribution today?}
\author{Robert J. Scherrer}
\affiliation{Department of Physics and Astronomy, Vanderbilt University, Nashville, TN
~~37235}

\begin{abstract}
Inhomogeneous big bang nucleosynthesis (BBN) produces a spatially inhomogeneous distribution of element abundances at $T \sim 10^9$ K, but subsequent element diffusion will tend to erase these inhomogeneities. We calculate the cosmological comoving diffusion
length for the BBN elements. This diffusion length is limited by atomic scattering
and is therefore dominated by diffusion when the atoms are neutral,
between the redshifts of recombination and reionization.  We find that the comoving
diffusion length today is $d_{com} \approx 70$ pc for all
of the elements of interest except $^7$Li, for which $d_{com}$ is an order of magnitude smaller because $^7$Li remains
ionized throughout the relevant epoch.  This comoving diffusion length
corresponds to a substellar baryonic mass scale and is roughly equal to the horizon scale at BBN.
These results lend support to the possibility that inhomogeneities
on scales larger than the horizon at BBN could
lead to a spatially inhomogeneous distribution of elements today, while purely
subhorizon fluctuations at BBN can result only in a homogeneous element distribution at present.

\end{abstract}

\maketitle

Since its development in the 1960s \cite{Hoyle,Peebles1,Peebles2,WFH} big bang nucleosynthesis (BBN) has become one of the main pillars
of modern cosmology.  While the baryon-photon ratio, $\eta$,
was once treated as the main quantity to be determined
by comparing the predictions of BBN to the observed
element abundances, the independent measurement
of $\eta$ from CMB observations has left the theory
of BBN with no free parameters.  Given the allowed
range for $\eta$ derived from the CMB, the predicted
BBN abundances of deuterium and $^{4}$He are
in excellent agreement with the observations. However, the predicted $^{7}$Li
abundance remains a factor of three larger than the abundance
inferred from observations; this discrepancy remains
unexplained at present \cite{lithium}.
For a recent review of BBN,
see Ref. \cite{Fields}.

Despite the successes of BBN, many modifications
to the theory have been proposed.
Perhaps the most widely investigated of these modified theories are models involving some sort of inhomogeneity at the epoch of BBN.  The most basic of such models postulate
isocurvature fluctuations, for which $\eta$ varies from
one region of the Universe to another \cite
{Wagoner,Epstein,Barrow,Yang,Copi,Leonard,BS}.  Most of these
investigations simply calculate the element abundances
as a function of $\eta$ in separate domains and average
the final results.  However,
an important effect arises when the length scale of the
fluctuations is longer than the proton diffusion length but shorter than the
neutron diffusion length; in this case neutron diffusion from high-density
to low-density regions will strongly modify the neutron-proton ratio.
These models were initially inspired by the possibility
of such density fluctuations arising naturally in the QCD phase transition \cite {Applegate,Alcock}, and this line of
investigation yielded an extensive literature
(see, e.g., Ref. \cite{Lara} and references therein).
Furthermore, several authors have pointed out that small regions
with
very large values of $\eta$ could lead to the primordial
production of elements much heavier than in standard BBN, including
even $r$-process elements \cite{AHS2,J1,Moriya,J2,J3,Arbey}.

However, spatial variations in $\eta$ are by no means the only inhomogeneous BBN models that have been considered.  Various authors have examined BBN with inhomogeneous
variations in the cosmological shear \cite{Olson},
inhomogeneous neutrino chemical potentials \cite{Dolgov,Whitmire,Stirling}, inhomogeneous
values of the baryon/dark matter ratio \cite{HNE},
inhomogeneous primordial magnetic fields \cite{Luo},
and inhomogeneous fluctuations in the dark radiation density \cite{Adshead}.
Models with small-scale antimatter domains have also been
discussed \cite{combes,Rehm1,Kurki1,Kurki2,Rehm2}.

Most of these studies of inhomogeneous BBN tacitly assume a uniform distribution of elements at present.  This assumption requires that post-BBN diffusion homogenizes the originally inhomogeneous distribution of elements.  However, a few authors
have considered the opposite case, in which the final distribution of elements
remains inhomogeneous today.  Models of this kind have sometimes been invoked to
explain anomalies in the observed element abundances.  Alternately,
observational constraints on the present-day inhomogeneity
of the primordial elements can be used to constrain such models.  Papers
that assume that inhomogeneities in BBN are preserved today include
Refs. \cite{Moriya,HNE,Adshead,Arbey}.

This dichotomy of approaches leads to an obvious question:
are the spatial inhomogeneities produced in inhomogeneous BBN preserved at present, or are they erased by subsequent element diffusion?  Clearly, inhomogeneities will be erased on sufficiently small scales and preserved on sufficiently large scales, so we can define a critical comoving length, $d_{com}$, that separates these two regimes.  The purpose of this paper is to determine $d_{com}$ and thereby cast some light on the validity of past assumptions regarding the final outcome of inhomogeneous BBN.

Surprisingly little work has been done on the question of post-BBN element diffusion in the early universe.  Moriya and Shigeyama \cite{Moriya} examined the particular case of helium diffusion on mass scales of order $10^6 ~M_\odot$ and concluded that diffusion would be insufficient for helium to diffuse out of such regions on a cosmological timescale. Several authors have investigated post-BBN element diffusion for the case of standard (homogeneous) BBN.  Because the nuclei have different masses, they will diffuse differently in gravitational potential wells, possibly leading to a small degree of inhomogeneity in the final element abundances.  Medvigy and Loeb \cite{Loeb} were the first to investigate this effect and concluded that the resulting inhomogeneities in the final abundances are very small, less than $0.1\%$.  Pospelov and Afshordi \cite {Pospelov} revisited the question of lithium diffusion, taking into account the fact that most lithium remains ionized after recombination. Kusakabe and Kawasaki investigated the differential diffusion of ionized lithium due to primordial magnetic fields \cite{Kusakabe1,Kusakabe2}. Medvedev et al. \cite{Medvedev} examined element diffusion during the formation
of the first galaxies. While we are, in a sense, performing the opposite calculation, we can incorporate many of the ideas from these papers in the present discussion.

We will assume an initially inhomogeneous distribution of nuclei at the epoch of BBN ($T \sim 10^9$ K) and examine the subsequent diffusion of these nuclei.  The nuclei produced
by BBN are $^1$H,
$^2$H, $^3$He, $^4$He, and $^7$Li.  (Note that $^3$He is
rarely examined as a probe of BBN because it is difficult to estimate
its primordial abundance from present-day observations, but we include it here
for completeness).  Furthermore, while heavier nuclei are not produced in standard BBN, they can be produced in some inhomogeneous models, so we will consider these nuclei as well.

As it diffuses, a given nucleus will undergo a random walk with mean free path $\lambda = 1/n \sigma$, where $n$ is the number density of targets off of which it can scatter, and $\sigma$ is the scattering cross section. The total distance $d$ that the nucleus diffuses at time $t$ is
$d = \sqrt {\lambda v t}$, so the diffusion length at time $t$
is given by
\begin{equation}
\label{d}
d = \sqrt{\frac{vt}{n\sigma}},
\end{equation}
while the comoving length scale today corresponding to this
diffusion length is
\begin{equation}
\label{dcom}
d_{com} = (1+z)\sqrt{\frac{vt}{n\sigma}}.
\end{equation}
These expressions assume that the
momentum transfer with each collision is a significant fraction
of the momentum of the nucleus.  As noted by Medvigy and Loeb \cite{Loeb},
the primary processes limiting element diffusion are collisions between the various
atomic species, so this assumption is justified. Because the scattering cross
section for ionized atoms is much larger than that between neutral atoms,
diffusion occurs most efficiently in the period after recombination ($1+z_{rec} \sim 1000$) and
before reionization ($1+z_{reion} \sim 10$).  Hence, we will consider only
diffusion in this redshift range. This allows us to ignore the effects of structure formation, since this redshift range lies within the linear regime, but we will reconsider these effects later.

Now
consider how these diffusion lengths scale with redshift.
For a given nucleus or atom with mass number $A$, the velocity in Eq. (\ref{d}) is given approxmately by
\begin{equation}
\label{velocity}
v \approx \sqrt{\frac{3kT_B}{m_p A}},
\end{equation}
where $T_B$ is the baryon temperature and $m_p$ is the proton mass.  After recombination, the baryon temperature is determined by a combination of
the residual free electron coupling to the background radiation, which tends
to keep the baryon temperature equal to the radiation temperature,
and adiabatic expansion, which drives $T_B$ to evolve as $(1+z)^2$
\cite{Peebles}.  Near the beginning of our epoch of interest (i.e., near recombination), we have $T_B = 2.7 (1+z)$ K, while close to the epoch of reionization,
the baryon temperature is well approximated as
$T_B = 0.02 (1+z)^2$ K \cite{Loeb,ScottMoss}.

The universe is matter-dominated during our epoch of interest, so
the time $t$ in equation (\ref{d}) is given by the standard
expression $t = 2.06 \times 10^{17} (\Omega_M h^2)^{-1/2} (1+z)^{-3/2}$ sec. (We
ignore the small contribution from $\Lambda$ near reionization, as well as the density
contribution from radiation near recombination).

In standard BBN, the baryonic matter at $z_{rec} > z > z_{reion}$ consists
almost entirely
of neutral helium atoms and neutral hydrogen atoms, with mass fractions of
$Y_p \approx 0.25$ and $1-Y_p \approx 0.75$, respectively.  This translates into atomic number densities of $n_{He} = (Y_p/4) n_B$ and $n_{H} = (1-Y_p)n_B$, where $n_B$ is
the baryon number density, given by
$n_B = 2.5 \times 10^{-7} (1+z)^3$ cm$^{-3}$.  It is the scattering off of these neutral hydrogen and helium atoms that determines the diffusion length of
each of the atomic species of interest.

Substituting all of these expressions into Eqs. (\ref{d}) and (\ref{dcom}), we find
that the diffusion length at redshift $z$ scales as $T_B^{1/4} (1+z)^{-9/4}$, corresponding to a comoving diffusion length
scaling as $T_B^{1/4} (1+z)^{-5/4}$.  Regardless of whether
$T_B$ scales as $1+z$ (near recombination) or $(1+z)^2$ (near reionization), we see that the comoving diffusion length increases
with decreasing redshift.  Thus, the diffusion length will be dominated
by its value at reionization.  This allows us to ignore the details
of recombination (except for nuclei which never recombine; see below) and concentrate on the reionization era.

Consider first the comoving diffusion length $d_{com}$ for
the isotopes of helium and hydrogen.  From Eq. (\ref{dcom}) we derive
\begin{eqnarray}
d_{com}(^1H) &=&(1+z)(3kT_B/m_p)^{1/4} t^{1/2}/(n_H \sigma_{HH} + n_{He}\sigma_{HeH})^{1/2}, \\
d_{com}(^2H) &=&(1+z)(3kT_B/2m_p)^{1/4} t^{1/2}/(n_H \sigma_{HH} + n_{He}\sigma_{HeH})^{1/2}, \\
d_{com}(^3He) &=&(1+z)(3kT_B/3m_p)^{1/4} t^{1/2}/(n_H \sigma_{HeH} + n_{He}\sigma_{HeHe})^{1/2}, \\
d_{com}(^4He) &=&(1+z)(3kT_B/4m_p)^{1/4} t^{1/2}/(n_H \sigma_{HeH} + n_{He}\sigma_{HeHe})^{1/2}.
\end{eqnarray}
where $\sigma_{AB}$ is the cross section for
scattering between neutral atoms $A$ and $B$.  The relevant
cross sections are taken from Ref. \cite{Medvedev}:
\begin{eqnarray}
\sigma_{HH} &=& 5.4 \times 10^{-15} {\rm cm}^2,\\
\sigma_{HeH} &=& 2.8 \times 10^{-15} {\rm cm}^2,\\
\sigma_{HeHe} &=& 5.4 \times 10^{-15} {\rm cm}^2.
\end{eqnarray}
Using our expressions for $T_B$, $t$, $n_H$ and $n_{He}$ , with
$Y_p = 0.25$, $\Omega_M = 0.3$ and $h= 0.7$, we find
that $d_{com}$ is roughly the same for all four nuclei of interest:  
\begin{equation}
d_{com} \approx 1 \times 10^{21} (1+z)^{-3/4} {\rm cm}.
\end{equation}
Note that scattering off of hydrogen dominates scattering off of helium for all of the elements
of interest because of the much larger number
density of hydrogen atoms.  Neglecting helium
scattering
changes the value of $d_{com}$ by less than
10\%, which is unimportant at the rough level of accuracy at which we are working here.

As noted, $d_{com}$ increases as $z$ decreases, so we 
set $z$ equal to the redshift of reionization to derive
the maximum comoving diffusion length.  Reionization is not an instantaneous process, so the redshift at which it occurs is somewhat ill-defined.  Here we will adopt the Planck value, $z_{reion} = 7.67 \pm 0.73$ \cite{Planck}.  As in Ref. \cite{Planck}, we will assume that the first reionization of helium occurs at the same redshift as the reionization of hydrogen,
while the later redshift at which
helium becomes fully reionized is irrelevant in this
calculation.  Then the comoving diffusion length for hydrogen, deuterium, and both isotopes of helium is
\begin{equation}
\label{dcom1}
d_{com} \approx 2 \times 10^{20} {\rm cm} \approx 70~ {\rm pc}.
\end{equation}
In regions that have undergone gravitational collapse (e.g., galaxies), this comoving length is not particularly meaningful
at the present. A more useful quantity is the baryonic mass contained inside a sphere of radius $d_{com}$; this
mass is
\begin{equation}
M_{com} \approx 7 \times 10^{-3} M_\odot.
\end{equation}
Thus, diffusion is insufficient to erase primordial inhomogeneities even on stellar length scales.

The other product of standard BBN, $^7$Li, must be dealt with separately.  Depending on the value of $\eta$, $^7$Li can be produced directly in BBN, or indirectly as the decay product of $^7$Be.  However,
the primordial $^7$Be decays to $^7$Li via electron capture at $z \sim 30,000$ \cite{Khatri}.  As this is well before our epoch of interest, we can treat all of the mass-7 nuclei as $^7$Li.

Because the ionization energy of $^7$Li is so small, nonthermal
UV photons from hydrogen recombination are sufficient
to keep $^7$Li in the singly-ionized state
throughout the epoch of interest \cite{Switzer}.
Then the relevant cross-section for lithium diffusion is the scattering cross section between ionized lithium and neutral hydrogen and helium.  This is given by \cite{Medvedev}
\begin{equation}
\sigma = 2.03 \times 10^{-15} (p_t/\AA^3)^{1/2}(E_{CM}/{\rm eV})^{-1/2} {\rm cm}^2,
\end{equation}
where $p_t$ is the polarizability of the neutral atom ($p_t = 
0.667 \AA^3$ for hydrogen and $p_t = 0.207 \AA^3$ for helium
\cite{Osterbrock}), and $E_{CM}$ is the center of mass collision energy.  Once again, $d_{com}$ increases with
decreasing redshift, reaching a maximum at $z_{reion}$, after which scattering between ionized lithium and ionized hydrogen and helium sharply decreases the diffusion length.  We then
find, for $^7$Li,
\begin{equation}
\label{dcom2}
d_{com} \approx 3 \times 10^{19} {\rm cm},
\end{equation}
or roughly an order of magnitude smaller than the comoving
diffusion length for the isotopes of hydrogen and helium.
This is because the cross section for ionized lithium
scattering off of the background neutral hydrogen and helium
at $z_{reion}$ is much larger than the corresponding
cross section for scattering between two neutral atoms.  As is the case
for the other
elements of interest, the diffusion of $^7$Li is limited primarily by scattering
off of hydrogen, with the additional scattering off of helium having only a very
small effect.

Elements heavier than lithium are not produced in significant quantities in standard BBN, but they can be produced in various inhomogeneous scenarios \cite{AHS2,J1,Moriya,J2,J3,Arbey}.  The diffusion length for heavier elements will depend on whether they fully recombine before reionization or if, like $^7$Li, they remain at least singly ionized at all times.
This, in turn, depends on the ionization energy of a given element. The ionization energies of the heavier elements vary widely, so either type of evolution is possible.
The continued ionization of lithium noted in Ref. \cite{Switzer} is driven by photons produced in the $n=2 \rightarrow 1$
transition in hydrogen, with $E_\gamma = 10.2$ eV, so we would expect elements
with ionization energies larger than 10 eV to
recombine before reionization, while the derivation of the
ionization history of elements with smaller ionization
energies requires a detailed calculation
as in Ref. \cite{Switzer}.
Further, the comoving diffusion lengths scale as $A^{-1/4}$ due to the smaller thermal
velocities of the heavier elements.  Thus, we expect Eq. (\ref{dcom1}) to provide
a good upper bound on the comoving diffusion length for any heavy elements produced
in inhomogeneous BBN. 

Now consider how these results affect inhomogeneous scenarios for BBN.  The comoving lengths in Eqs. (\ref{dcom1}) and
(\ref{dcom2}) are of the order of the horizon size
at $T\sim 10^{10}$ K, when BBN begins with the freeze-out of the $n \leftrightarrow p$ reactions.  If inhomogeneities
are produced by a purely causal mechanism in such a way
that there are no inhomogeneities on scales larger than the horizon at BBN, then the resulting inhomogeneous
element distribution will be completely erased by subsequent
element diffusion, resulting in a homogeneous distribution
of elements at the present (albeit different from the abundances produced by standard BBN).  On the other hand, inhomogeneities on scales larger than the horizon at BBN would exceed the comoving element diffusion length and would therefore
result in an inhomogeneous final distribution of elements.

Although these diffusion lengths suggest that BBN inhomogeneities could be preserved on stellar length scales,
we expect further mixing to occur due to turbulence during the process of galaxy formation.  The exact
details of this mixing, however, remain somewhat unresolved (see, e.g., Ref. \cite{turbulence} for a recent discussion).  Indeed, Adshead et al. \cite{Adshead} have suggested that
the appropriate length scale for primordial element inhomogeneities to
be erased is on the scale of individual galaxies, with the
inhomogeneities preserved between different galaxies.
While galaxy
formation (along with subsequent chemical evolution) will
certainly smooth out primordial inhomogeneities
on scales larger than those given by Eqs. (\ref{dcom1}) and (\ref{dcom2}), it seems unlikely that they would erase all primordial inhomogeneities on the scale of an entire galaxy.
Hence, our results add support to the approach taken in
Refs. \cite{Moriya,HNE,Adshead,Arbey}.  Unless inhomogeneities
in BBN are taken to be sufficiently small (completely sub-horizon at BBN), one cannot assume that the present-day element abundances are simply a homogeneous average of the originally inhomogenous distribution of elements.
	
It is intriguing that our results yield a diffusion length for
$^7$Li much smaller than for the other primordial elements,
given the longstanding primordial lithium problem \cite{lithium}.  In principle, one could devise a model with a fine-tuned inhomogeneity scale such that the spatial distribution of $^7$Li
was inhomogeneous today, while the other elements were homogeneously distributed.  However, it is not at all clear how
such a model could provide a satisfactory solution to the lithium problem. 
	
Our calculation here is admittedly rough; a more accurate calculation would require an integration of the diffusion equation for the various elements of interest.  However, we would not expect our results to differ from an exact calculation by factors of more than order unity.

\begin{acknowledgments}

R.J.S. was supported in part by the Department of Energy (DE-SC0019207).  He thanks A. Loeb for helpful comments.

\end{acknowledgments}

\end{document}